# Exponential Signal Reconstruction with Deep Hankel Matrix Factorization

Yihui Huang[†], Jinkui Zhao[†], Zi Wang, Vladislav Orekhov, Di Guo, Xiaobo Qu[*]

*Abstract*—Exponential function is a basic form of temporal signals, and how to fast acquire this signal is one of the fundamental problems and frontiers in signal processing. To achieve this goal, partial data may be acquired but result in severe artifacts in its spectrum, which is the Fourier transform of exponentials. Thus, reliable spectrum reconstruction is highly expected in the fast data acquisition in many applications, such as chemistry, biology, and medical imaging. In this work, we propose a deep learning method whose neural network structure is designed by imitating the iterative process in the model-based state-of-the-art exponentials reconstruction method with low-rank Hankel matrix factorization. With the experiments on synthetic data and realistic biological magnetic resonance signals, we demonstrate that the new method yields much lower reconstruction errors and preserves the low-intensity signals much better than compared methods.

## I. INTRODUCTION

EXPONENTIAL function is a basic signal form in signal processing. In many practical applications, signals can be approximated by the superposition of a few exponential functions. Examples include antenna signals in telecommunication [1-3], images in fluorescence microscopy [4], analog-to-digital conversion in electronic systems [5], signal functions in the theory of the finite rate of innovation [6], planar polygon recovering[7], and time-domain signals in biological Nuclear Magnetic Resonance (NMR) spectroscopy [8-14]. Thus, achieving high-quality recovery of the exponential signals has great significance. The signal of interest is modeled as the sum of exponentials as follows:

$$x(n\Delta t) = \sum_{j=1}^{J} (A_j e^{i\phi_j}) e^{-\frac{n\Delta t}{\tau_j}} e^{i2\pi f_j n\Delta t}, \quad (1)$$

where $f_j \in [0,1)$ is the normalized frequency, $\phi_j \in [0, 2\pi)$ is the phase, $A_j \in \mathbb{R}_+$ is the amplitude, $\tau_j \in \mathbb{R}_+$ is the damping factor, $J$ is the number of exponentials, $j=1,...,J$, and $\Delta t$ is the data acquisition (also called sampling) interval, $n$ is the index of fully sampled data points in the time domain with $n=1,2,\cdots,N$. Accordingly, the sampled data can be represented with a vector $\mathbf{x} = [x_1,...,x_N]^T \in \mathbb{C}^N$. By performing the Fourier transform $\mathcal{F}$ on $\mathbf{x}$, one can obtain a full spectrum $\mathcal{F}\mathbf{x}$.

Reconstruction from undersampled or corrupted signal [15] and fast data acquisition of exponentials is one of the frontier and important problems in signal processing [6, 8, 10-12, 16-19]. To enable fast acquisition, the main approach is to acquire the non-uniformly sampled time-domain data $\mathbf{y}=\mathcal{U}\mathbf{x}$ where $\mathcal{U} \in \mathbb{R}^{M \times N}$ denotes the undersampling operator with $M < N$. As the system is underdetermined, one has to introduce some priors to recover the signal $\mathbf{x}$ from $\mathbf{y}$. For instance, in biological NMR spectroscopy [8], exponential signal reconstruction from the undersampled data has enabled the acceleration factor of 4 to 10 in fast data acquisition, which significantly reduces the data acquisition time [8, 20-25].

To reconstruct exponentials $\mathbf{x}$ from the undersampled data $\mathbf{y}$, there are two main model-based approaches, which explore the sparsity of $\mathcal{F}\mathbf{x}$ in the frequency domain and low rankness of $\mathbf{x}$ in the time domain. Without particularly considering the damping factor, i.e., the signal decay, the former has been evidenced powerful in conventional compressed sensing [26] and state-of-the-art total variation or atomic norm methods [27].

The sparsity of the spectrum $\mathcal{F}\mathbf{x}$, however, may not be satisfied for a fast decaying exponential, as broad spectral peaks (Fig. 1) may be introduced with a damping factor [8]. Thus, researchers exploit the mathematical property of the signal in the time domain, i.e. using the presentation of $\mathbf{x}$ in the form of Hankel matrix[8, 12, 19, 28] or higher-order tensor[11, 13, 29]. When $\mathbf{x}$ is converted into a Hankel matrix, the matrix rank is equal to the number of underlying exponentials[8]. Thus, the total rank of Hankel matrix is low, if the number of exponentials is much smaller than the number of data points $N$ [8, 17, 18, 30-33]. Then, $\mathbf{x}$ can be recovered by enforcing the low rankness of

This work was supported in part by the National Natural Science Foundation of China (62122064, 61971361, 61871341, and 61811530021), the National Key R&D Program of China (2017YFC0108703), Natural Science Foundation of Fujian Province of China (2021J011184), and Xiamen University Nanqiang Outstanding Talents Program. ([†]Equal Contributions, [*] Corresponding author)

Yihui Huang, Jinkui Zhao, Zi Wang, and Xiaobo Qu are with the Department of Electronic Science, Biomedical Intelligent Cloud R&D Center, National Institute for Data Science in Health and Medicine, Xiamen University, Xiamen, China (e-mail: quxiaobo@xmu.edu.cn).

Di Guo is with the School of Computer and Information Engineering, Xiamen University of Technology, Xiamen, China.

Vladislav Orekhov is with the Department of Chemistry and Molecular Biology, University of Gothenburg, Gothenburg, Sweden.

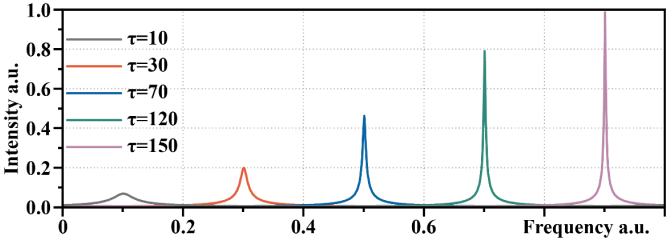

Fig. 1. A synthetic spectrum with low-intensity peaks that are introduced by small damping factors. Note: Five peaks are generated with the same amplitudes and phases in (1), and frequencies are set from 0.1 to 0.9. The corresponding damping factors for peaks from left to right are 10, 30, 70, 120 and 150, respectively. Thus, smaller damping factor leads to broader and lower peaks.

the Hankel matrix [8]. This approach is called Low-Rank Hankel Matrix Completion (LRHMC) [8, 17, 18, 30-33].

Finding a fast algorithm remains challenging for LRHMC. Since the rank minimization problem is NP-hard, some approaches utilize the nuclear norm to relax it into a convex optimization problem [8, 18, 32, 34]. These approaches lead to the faithful reconstruction of exponentials and high fidelity spectra in practical applications, e.g. fast biological NMR [8]. However, minimization of the nuclear norm commonly utilizes time-consuming Singular Value Decomposition (SVD) [35], which leads to excessively lengthy calculations [8]. Although matrix factorizations [10, 12, 36, 37] were proposed to avoid SVD, the computational time is still long due to many iterations in these model-based optimization algorithms.

Inspired by the prominent achievements in the field of deep learning (DL)[38, 39], a proof-of-concept of exponentials reconstruction with DL was proposed and applied to fast NMR with undersampling [16]. The approach offers ultrafast computation as the traditional iterative reconstruction is replaced by a forward process once the network has been trained on graphics processing units (GPUs) with parallel computing. Furthermore, it was demonstrated that training of the neural network can be achieved using solely synthetic exponential signals, which lifted the prohibiting demand for a large volume of realistic biological training data that are expensive and difficult to obtain [16, 40].

However, spectra reconstructions obtained by the original DL approach [16] often show distortions for low-intensity exponentials (and spectral peaks). Besides, the trained DL model represents a black box in the sense that it lacks the interpretability of the neural network structure and thus does not provide guidance for its further improvement. A possible solution for this problem is to base the network design on a well-established model-based optimization method [39, 41-44].

In this work, we propose a DL neural network for recovering exponentials by imitating the established model-based low-rank Hankel reconstruction algorithm [8, 10]. Several deep learning methods based on low-rank[45, 46], non-negative[47, 48] and low-rank plus sparse constraints[49, 50] are introduced. These methods use shallow linear networks or deep nonlinear networks. The latter requires non-negative matrices or uses time-consuming SVD, and thus is not suitable for the reconstruction of exponentials with large sizes.

To reconstruct exponential signals without SVD, we introduce Deep Hankel Matrix Factorization network (DHMF). The design of our network starts from constraining the low-rank property of the Hankel matrix which is arranged from exponential signals and adopts matrix factorization-based LRHMC [8, 10] which are SVD-free. Given the success of the training network using solely synthetic data [16], we also synthesize exponentials and then feed into the network to learn the mapping from the input, i.e. undersampled exponentials, to the output, fully sampled exponentials.

Our main contributions are summarized as follows: 1) The proposed DHMF provides a new direction for low-rank approximation with deep learning matrix factorization and demonstrates the network interpretation by analyzing the singular value of reconstruction results. 2) Compared with the state-of-the-art model-based LRHM, LRHMF, and DL-based methods, DHMF is much faster than the former and possesses a more understandable network architecture than the latter. 3) DHMF achieves lower reconstruction error than the compared methods on both synthetic and realistic NMR data, and can preserve the low-intensity signals much better.

The rest of this paper is organized as follows. Sections II and III introduce mathematical notations and model-based methods, respectively. In Section IV, we propose the method and provide the interpretation of the network structure. Section V elaborates experiments setup and analyzes the results. Section VI and VII provides discussions and conclusions.

## II. NOTATIONS

We start with the notation used throughout this paper. Scalar $x$, vector $\mathbf{x}$, matrix $\mathbf{X}$, and identity matrix $\mathbf{I}$. The $j$-th entry of a vector is denoted by $\mathbf{x}_j$ and the $(i, j)$-th entry of a matrix is $\mathbf{X}_{ij}$. The notation $\|\cdot\|_p$ represents the standard $p$-norm, $\|\cdot\|_*$ nuclear norm, and $\|\cdot\|_F$ Frobenius norm. The transpose of vector and matrix are denoted by $\mathbf{x}^T$ and $\mathbf{X}^T$, while their conjugate transpose is donated by $\mathbf{x}^H$ and $\mathbf{X}^H$. $\mathbf{X}^{-1}$ represents the inverse of the matrix $\mathbf{X}$. Superscript $k$ denotes the number of iterations.

Operators are denoted by calligraphic letters. $\mathcal{R}$ denotes the Hankel operator mapping the vector $\mathbf{x} \in \mathbb{C}^N$ to a Hankel matrix $\mathcal{R}\mathbf{x} \in \mathbb{C}^{N_1 \times N_2}$ with $N_1 + N_2 = N + 1$ as follows:

$$\mathcal{R}: \mathbf{x} \in \mathbb{C}^{N_1+N_2-1} \mapsto \mathbf{X} \in \mathbb{C}^{N_1 \times N_2}, \mathbf{X}_{ij} = \mathbf{x}_{i+j-1}, \quad (2)$$

Corresponding inverse operator $\mathcal{R}^*$, which turns a Hankel matrix to a vector and satisfies $\mathcal{R}^*(\mathcal{R}\mathbf{x}) = \mathbf{x}$, is given by

$$\mathcal{R}^*: \mathbf{X} \in \mathbb{C}^{N_1 \times N_2} \mapsto \mathbf{x} \in \mathbb{C}^{N_1+N_2-1}, \mathbf{x}_g = \frac{\sum_{i+j-1=g} \mathbf{X}_{ij}}{\sum_{i+j-1=g} (\mathcal{R}\mathbf{o})_{ij}}, \quad (3)$$

where $\mathbf{o} \in \mathbb{C}^{N_1+N_2-1}$ is the vector whose elements are all ones for any $i \in \{1,...,N_1\}$ and $j \in \{1,...,N_2\}$ and the sum and Hankel operator $\sum_{i+j-1=g} (\mathcal{R}\mathbf{o})_{ij}$ computes the number of entries in the anti-diagonal direction of the matrix.

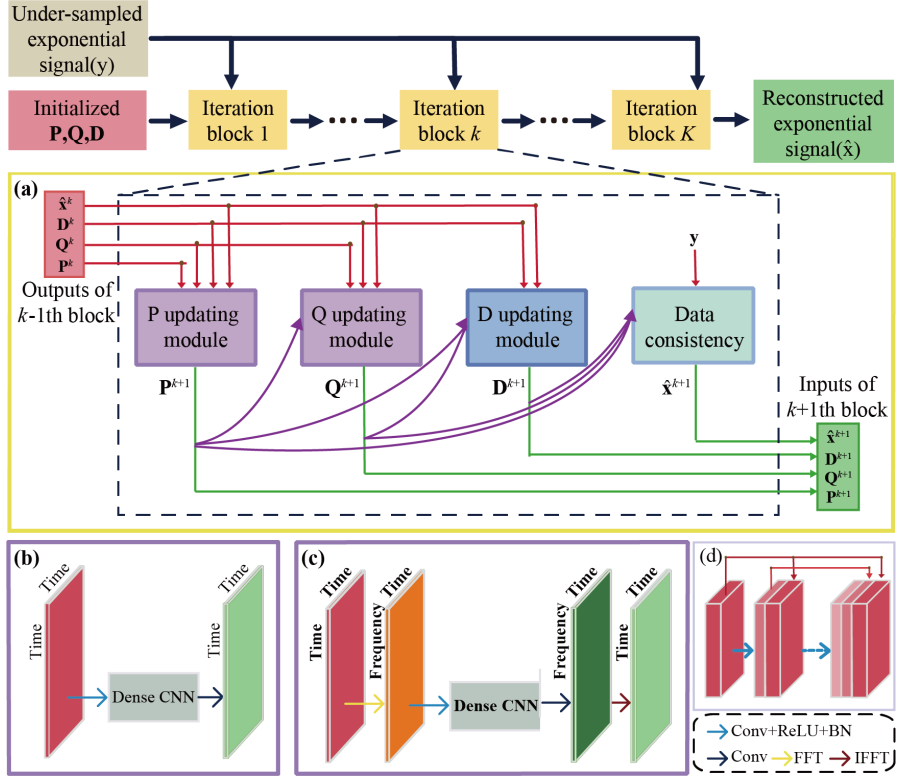

Fig. 2. The architecture of DHMF. (a) the general process of the *k*-th block, (b) P/Q modules with time domain convolution in the basic DHMF, (c) P/Q modules with frequency domain convolution in the enhanced DHMF, (d) dense convolutional neural network.

## III. RELATED WORK

In this section, we briefly review two typical model-based reconstruction methods, including the Low-Rank Hankel Matrix (LRHM) [8] and Low-Rank Hankel Matrix Factorization (LRHMF) [10] and one state-of-the-art deep learning method [16].

LRHM utilizes the low-rank property of the Hankel matrix $\mathcal{R}\mathbf{x}$ to model the reconstruction problem as

$$\min_{\mathbf{x}} \|\mathcal{R}\mathbf{x}\|_* + \frac{\lambda}{2}\|\mathbf{y} - \mathcal{U}\mathbf{x}\|_2^2, \quad (4)$$

where $\mathcal{U}$ denotes the undersampling operator, $\|\mathcal{R}\mathbf{x}\|_*$ is the nuclear norm, defined as the sum of singular values, as a proxy of the matrix rank, $\|\mathbf{y} - \mathcal{U}\mathbf{x}\|_2^2$ is the data consistency term, and the regularization parameter $\lambda$ tradeoffs between the low rankness and data consistency. An Alternating Direction Method of Multipliers (ADMM)[52] method was introduced to minimize the objective function (4).

To avoid computing the singular values that usually requires time-consuming SVD, the LRHMF employs a factorization $\mathcal{R}\mathbf{x} = \mathbf{P}\mathbf{Q}^H$ to replace the nuclear norm[51]:

$$\|\mathbf{X}\|_* = \min_{\mathbf{P},\mathbf{Q}} \frac{1}{2}(\|\mathbf{P}\|_F^2 + \|\mathbf{Q}\|_F^2), \ s.t. \ \mathbf{P}\mathbf{Q}^H = \mathbf{X} \quad (5)$$

where $\mathbf{P} \in \mathbb{C}^{N_1 \times R}$ and $\mathbf{Q} \in \mathbb{C}^{N_2 \times R}$. Then, equation (4) is reformulated as:

$$\min_{\mathbf{x},\mathbf{P},\mathbf{Q}} \frac{1}{2}(\|\mathbf{P}\|_F^2 + \|\mathbf{Q}\|_F^2) + \frac{\lambda}{2}\|\mathbf{y} - \mathcal{U}\mathbf{x}\|_2^2, \ s.t. \ \mathcal{R}\mathbf{x} = \mathbf{P}\mathbf{Q}^H. \quad (6)$$

The augmented Lagrangian function of (6) is

$$L(\mathbf{x},\mathbf{P},\mathbf{Q},\mathbf{D}) = \frac{1}{2}\|\mathbf{P}\|_F^2 + \frac{1}{2}\|\mathbf{Q}\|_F^2 + \frac{\lambda}{2}\|\mathbf{y}-\mathcal{U}\mathbf{x}\|_2^2 \\ + \langle \mathbf{D}, \mathcal{R}\mathbf{x}-\mathbf{P}\mathbf{Q}^H \rangle + \frac{\beta}{2}\|\mathcal{R}\mathbf{x}-\mathbf{P}\mathbf{Q}^H\|_F^2 \quad (7)$$

where $\mathbf{D}$ denotes the augmented Lagrange multiplier, $\langle \cdot, \cdot \rangle$ represents an inner product operator, parameters $\lambda > 0$, and $\beta > 0$ is used to balance between each term. Similar to LRHM, ADMM [52] method was introduced to minimize (7) and the solution at the *k*-th iteration is:

$$\begin{cases} \mathbf{x}^{k+1} = (\lambda \mathcal{U}^*\mathcal{U} + \beta \mathcal{R}^*\mathcal{R})^{-1}(\lambda \mathcal{U}^*\mathbf{y} + \beta \mathcal{R}^*(\mathbf{P}^k(\mathbf{Q}^k)^H - \mathbf{D}^k)) \\ \mathbf{P}^{k+1} = \beta(\mathcal{R}\mathbf{x}^{k+1} + \mathbf{D}^k)\mathbf{Q}^k(\beta(\mathbf{Q}^k)^H\mathbf{Q}^k + \mathbf{I})^{-1} \\ \mathbf{Q}^{k+1} = \beta(\mathcal{R}\mathbf{x}^{k+1} + \mathbf{D}^k)^H\mathbf{P}^{k+1}(\beta(\mathbf{P}^{k+1})^H\mathbf{P}^{k+1} + \mathbf{I})^{-1} \\ \mathbf{D}^{k+1} = \mathbf{D}^k + \tau(\mathcal{R}\mathbf{x}^{k+1} - \mathbf{P}^{k+1}(\mathbf{Q}^{k+1})^H) \end{cases}, \quad (8)$$

where $\tau$ is a step size. LRHMF runs much faster and obtains comparable reconstruction errors than LRHM in application [10].

Recently, deep learning for reconstructing exponential NMR signals has been introduced[16]. This method stacks several dense convolution neural networks together and refines the intermediately reconstructed spectra by enforcing data consistency. The spectrum artifacts in the frequency domain are gradually removed through the data flow of the network. Throughout the paper, we call this approach Deep Learning NMR (DLNMR). The method, which uses a neural network trained on purely synthetic exponentials, reconstructs faithfully most of the spectra. However, a number of low-intensity peaks are distorted or even lost in the spectra. Thus, further improvement is needed.

## IV. PROPOSED METHOD

In this section, we propose the Deep Hankel Matrix Factorization (DHMF) neural network and introduce its enhanced version.

### A. Basic DHMF

Similar to the iterative process of LRHMF [10], the proposed DHMF consists of three updating modules $\mathbf{P}$, $\mathbf{Q}$, $\mathbf{D}$, and one data consistency module (Fig. 2). Each module corresponds to the updating of four intermediate variables in (8). The four modules constitute a block and the whole network is a stack of $K$ blocks.

#### 1) Updating Module of $\mathbf{P}$ and $\mathbf{Q}$

Here, the dense convolutional network [53], followed by Rectified Linear Unit (ReLU) and Batch Normalization (BN), is adopted to learn the updating operators $\mathcal{P}$ and $\mathcal{Q}$ because it enables maximum information flow in the network. To leave the historical information of $\mathbf{P}^k$ and $\mathbf{Q}^k$ transfer to the following layers, $\mathbf{P}$ and $\mathbf{Q}$ updating modules of $k$-th $(k=1,2...,K)$ block are designed as:

$$\begin{aligned}\mathbf{P}^{k+1} &= \mathcal{P}^k((\mathcal{R}\hat{\mathbf{x}}^k + \mathbf{D}^k)\mathbf{Q}^k, \mathbf{Q}^k, \mathbf{P}^k) \\ \mathbf{Q}^{k+1} &= \mathcal{Q}^k((\mathcal{R}\hat{\mathbf{x}}^k + \mathbf{D}^k)^H \mathbf{P}^{k+1}, \mathbf{P}^{k+1}, \mathbf{Q}^k)\end{aligned}, \quad (9)$$

In the implementation, the real and imaginary parts of the input are concatenated together and then separated into two parts of complex variables after updating.

#### 2) Updating Module of $\mathbf{D}$

The updating module of $\mathbf{D}$ is calculated by:

$$\mathbf{D}^{k+1} = \mathbf{D}^k + \tau(\mathcal{R}\hat{\mathbf{x}}^k - \mathbf{P}^{k+1}(\mathbf{Q}^{k+1})^H), \quad (10)$$

where $\tau$ is set as a constant and $\mathbf{D}$ initialized as a zero matrix. The $\mathbf{D}$ is not learnable due to suboptimal results (See discussions in Section VI. C.)

#### 3) Data Consistency Module

Since the time domain signal $\mathbf{x}$ in (8) should be aligned to acquired data [16, 54], the data consistency module is designed as:

$$\hat{\mathbf{x}}^{k+1} = \mathcal{S}(\mathbf{y}, \mathcal{R}^*(\mathbf{P}^{k+1}(\mathbf{Q}^{k+1})^H - \mathbf{D}^{k+1})), \quad (11)$$

where $\mathcal{S}$ denotes the data consistency operator. Let $\tilde{\mathbf{x}}^{k+1}$ be the restored data obtained by $\mathcal{R}^*(\mathbf{P}^{k+1}(\mathbf{Q}^{k+1})^H - \mathbf{D}^{k+1})$, equation (11) is equal to the following relationship:

$$\hat{\mathbf{x}}_n^{k+1} = \mathcal{S}(\mathbf{y}, \tilde{\mathbf{x}}_n^{k+1}) = \begin{cases} \tilde{\mathbf{x}}_n^{k+1}, & \text{if } n \notin \Omega \\ \dfrac{\tilde{\mathbf{x}}_n^{k+1} + \lambda \mathbf{y}_n}{1+\lambda}, & \text{if } n \in \Omega \end{cases}, \quad (12)$$

where $\Omega$ is the set of positions for sampled exponential signal, $n$ is the index of the signal. Equation (12) implies that the exponential signal at the location of sampled data points should maintain a tradeoff between the sampled data $\mathbf{y}$ and the restored data $\tilde{\mathbf{x}}_n^{k+1}$ under a parameter $\lambda$, which possesses the same meaning as that in (4).

Overall, time-domain convolution networks are trained in $\mathbf{P}$ and $\mathbf{Q}$ updating modules, while in $\mathbf{D}$ updating modules and data consistency modules, variables are directly calculated and no parameter needs to be trained. With the four modules in each block, spectrum artifacts are reduced by increasing the number of blocks, which is similar to the observation in DLNMR [16].

### B. Enhanced DHMF

In the basic DHMF, all the convolution is performed in the time domain (Fig. 2(b)). This operation, however, may lose some global information as the convolution kernel is usually not too long. To better explore this information, we first use the fast Fourier transform (FFT) or inverse FFT (IFFT) in the columns of variables $\mathbf{P}$ and $\mathbf{Q}$ and then perform the convolution in the frequency domain (Fig. 2(c)).

As a representative example in Fig. 3. The enhanced DHMF better preserves the low-intensity peaks while the basic DHMF still presents some artifacts. To further evaluate the reconstruction error, we use the Relative Least Normalized Error (RLNE) [11, 12, 19] defined as:

$$\text{RLNE} = \frac{\|\mathbf{x} - \hat{\mathbf{x}}\|_2}{\|\mathbf{x}\|_2}, \quad (13)$$

to measure the difference between the reconstruction $\hat{\mathbf{x}}$ and the noise-free fully sampled signal $\mathbf{x}$.

In the following parts, we only use enhanced DHMF and call it DHMF for simplicity.

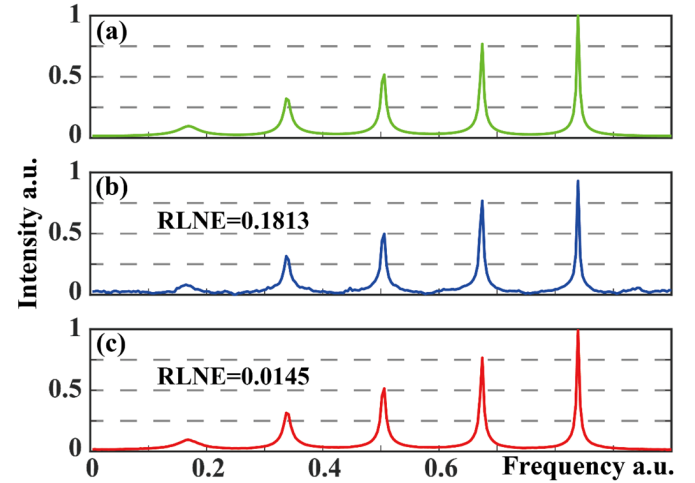

Fig. 3. Reconstructed spectra from undersampled five synthetic exponentials, (a) is the fully sampled signal, (b) and (c) are the reconstructed by basic DHMF and enhanced DHMF, respectively. Note: The length of the fully sampled data is 255 and 25% is acquired.

### C. Loss Function and Hyper-parameters

Following the most general strategy in deep learning, the sum of the mean squared errors (equation (14)) is chosen as the loss function to measure the difference between the reconstructed signal $\hat{\mathbf{x}}$ and noise-free fully sampled signal $\mathbf{x}$. Besides, to make the output of each block $\hat{\mathbf{x}}_q^k$ approaches the ground truth $\mathbf{x}$, the loss function on $\hat{\mathbf{x}}_q^k(\Theta) - \mathbf{x}_q$ is also constrained. This process was suggested to make the solution steadily approach the solution in previous work[16].

Besides, as the matrix factorization constraint $\mathcal{R}\mathbf{x}=\mathbf{PQ}^H$ in (6) should be satisfied in the conventional model-based method LRHMF, we also add the square of the $l_2$ norm of $(\mathcal{R}^*(\mathbf{PQ}^H) - \mathbf{x})$ in the loss function.

TABLE I
PARAMETERS FOR 1D SYNTHETIC EXPONENTIAL SIGNAL

| Randomly selected range | Parameters of exponentials | | | | |
|---|---|---|---|---|---|
| | Number of exponentials($J$) | Amplitude ($A$) | Normalized frequency($f$) | Damping factor($\tau$) | Phase ($\phi$) |
| Minimum | 1 | 0.05 | 0 | 10 | 0 |
| Precision | 1 | 2.2 x$10^{-16}$ | 2.2 x$10^{-16}$ | 2.8 x$10^{-14}$ | 8.8 x$10^{-16}$ |
| Maximum | 10 | 1 | 1 | 179.2 | $2\pi$ |

Note: The number of peaks increase from 1 to 10 and length of signal is 255, each subset contains a specific number of peaks and other parameters are acquired uniformly at random using Matlab. Thus, we have 10 subsets and each of them consists 4000 signals. The increment is acquired by calculation the accuracy of the float numbers.

Overall, the final loss function is

$$\mathcal{L}(\Theta) = \frac{1}{Q}\sum_{q=1}^{Q}\sum_{k=1}^{K}(\|\hat{\mathbf{x}}_q^k(\Theta) - \mathbf{x}_q\|_2^2 + \gamma \|\mathcal{R}^*(\mathbf{P}_q^k(\Theta)\mathbf{Q}_q^k(\Theta)^H) - \mathbf{x}_q\|_2^2)$$ (14)

where $\gamma$ is the regularization parameter. Subscript $q$ means the $q$-th index in training data of the total $Q$ sampling trail. $\Theta$ is the learnable network parameters including all the learnable weights of convolution layers and batch normalization layers in the **P** and **Q** updating module. In the implementation, optimal parameters of the network are obtained by minimizing (14) with Adam [55] optimizer.

The size R of variable **P** and **Q** is set to 20, which is twice as the maximum number 10 of exponential of signal in training dataset experimentally. For network structure, the kernel size is $3\times 3$, the number of the dense layers $L$ is 8, and the block number $K$ is 5. The parameter $\tau$ in **D** module, the $\lambda$ in the data consistency module, and the regularization parameter $\gamma$ are set to $10^{-3}$, $\sqrt{10}$ and $10^{-2}$, respectively. For network training, the learning rate starts from $10^{-3}$ and gradually reduces to $10^{-5.5}$ when the loss function is no longer significantly reduced.

*D. Network Interpretation*

To demonstrate the network interpretation of DHMF, we show the intermediate reconstructions and corresponding singular values of each iteration block (Fig. 4). As the data proceeds through the successive blocks, a low-rank solution is gradually obtained by DHMF since the rankness indicator, the nuclear norm of the Hankel matrix, decreases steadily (Fig. 4(g)) and finally becomes close to that of the fully sampled signal. The DLNMR, however, increases the nuclear norm at the final iteration block. In each block(Figs. 3(h)-(l)), the DHMF provides a much better approximation of singular values than DLNMR. At the last block(Fig. 4(l)), DHMF provides very close (although not the same) singular values to that of the fully sampled signal.

These observations imply that DHMF implicitly learns the approximation of low-rank property, and can provide a better interpretation of the reconstruction in the network. This phenomenon is mainly contributed to our network architecture: firstly, the proposed imitates the iteration process (8) of LRHMF[9] to minimize the nuclear norm of $\mathbf{PQ}^H$ through (5). Secondly, the size $R$ of variable **P** and **Q** in the proposed is set as 20, which is far less than that of Hankel matrix of the reconstructed signal. This design also implicitly forces reconstructed signals being low rank.

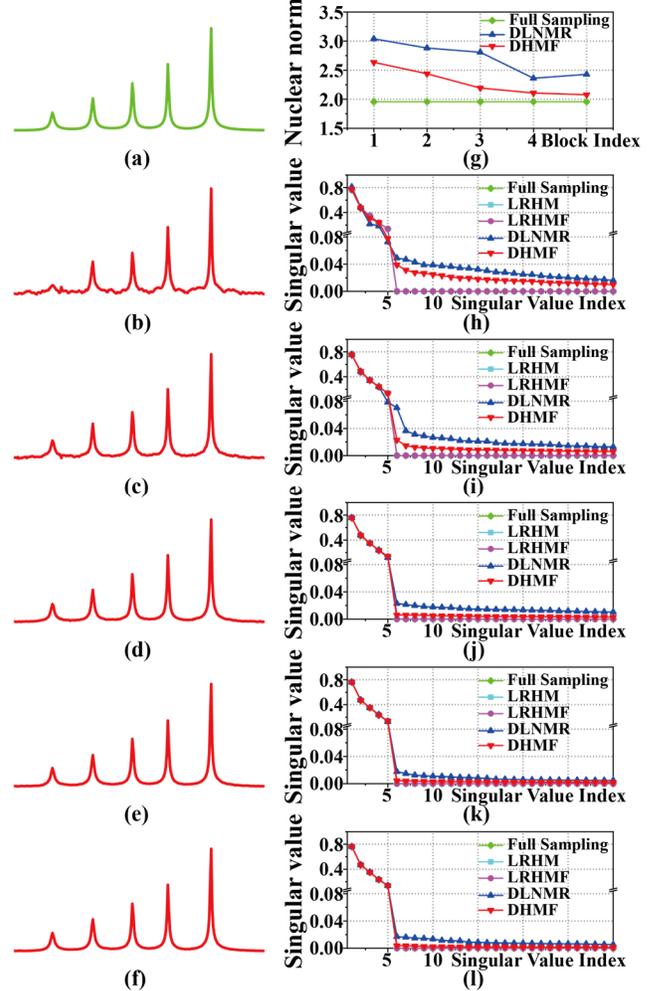

Fig. 4. The reconstructed spectra and singular values at each block. (a) Fully sampled spectrum, (b)-(f) the reconstructed spectrum by the 1st to 5th blocks, (g) the nuclear norm of Hankel matrix, and (h)-(l) denote corresponding singular values of the output of each block. Note: The singular values of full sampling and LRHM are very close thus the corresponding curves in (h)-(l) are nearly overlapped.

V. RESULTS

In this section, we evaluate the performance of the proposed DHMF on both synthetic exponential signals and realistic biological NMR spectroscopy with three state-of-the-art approaches, including LRHM [8], LRHMF [10], and DLNMR [16]. Both LRHM and LRHMF are typical model-based iterative algorithms, and LRHMF is an SVD-free version of LRHM. DLNMR is a state-of-the-art deep learning method for exponential signal reconstruction. The data consistency weight $\lambda$ is set to $\sqrt{10}$ in DHMF, $10^3$ in DLNMR, and $10^{2.5}$ in both LRHM and LRHM. All these parameters are optimized for the lowest reconstruction errors.

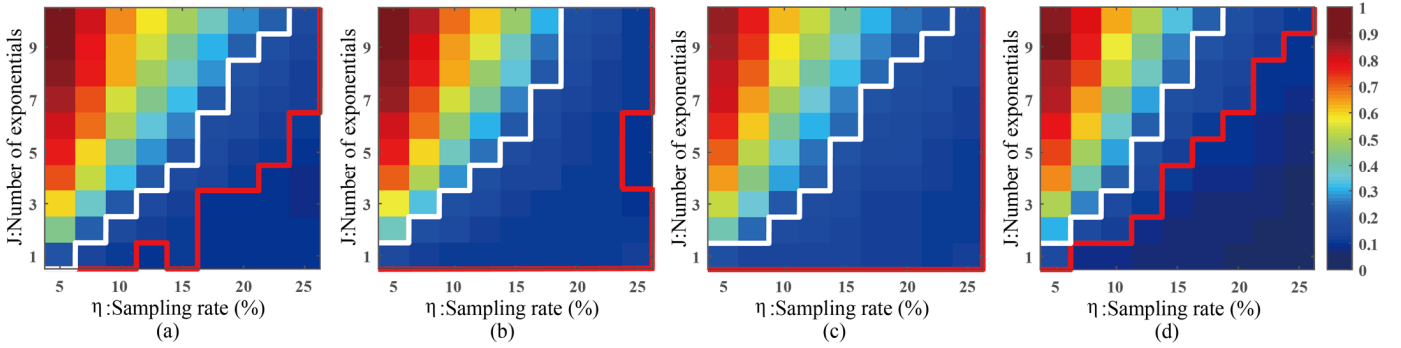

Fig. 5. Average reconstruction error, RLNE, for the synthetic signal. (a) LRHM, (b)LRHMF, (c) DLNMR, (d)DHMF. Note: Each color reflects the average RLNEs over 100 Monte Carlo trials of undersamplings. Red (or white) line indicates an empirical boundary where the threshold of reconstruction error, RLNE, is 0.1 (or 0.2). The reconstruction error in the upper region above (or lower region below) the boundary is greater (or lower) than the threshold. The DLNMR did not obtain any smaller reconstruction error than 0.1 thus the region below the red line is empty. All reconstruction errors are listed in Supplement TABLE S3-1 to S3-4.

All experiments were carried out in a computer server equipped with one Intel i9 9900X CPU (3.5 GHz, ten cores), 128 GB RAM, and three NVIDIA RTX 2080Ti GPU cards. The proposed DHMF and DLNMR networks were implemented in python 3.6, using the Keras 2.2.4 package and Tensorflow 1.14.0 as backend. Both LRHM and LRHMF coded in MATLAB (Mathworks Inc.) were parallelized to reduce the computation time under ten CPU cores maximally.

In the following, we use the sampling rate $\eta$, which is defined as $\eta=M/N$ where $M$ and $N$ are the numbers of partially sampled and fully sampled data points. For each fully sampled data points, a corresponding undersampling operator $\mathcal{U}$ is generated following the Poisson-gap sampling scheme[56].

### A. Training Data

The networks used for reconstructing both the synthetic and realistic data are trained on the same dataset, which solely consists of synthetic exponential signals generated following the parameters in TABLE I. Since the additive Gaussian noise is common in NMR spectroscopy [11, 12, 16, 19, 22, 57], the noisy training data is generated according to $\mathbf{y} = \mathcal{U}\mathbf{x} + \delta$ where $\delta$ is the Gaussian noise with a standard deviation of $5\times10^{-2}$ and $\mathbf{x}$ is the noise-free fully sampled exponential signals.

By feeding a signal pair $(\mathbf{y},\mathcal{U},\mathbf{x})$ into the network, the learnable parameters of DHMF are optimized until the learning rate decreases to the minimal value $10^{-5.5}$. In the training dataset, a total of $Q=40000$ signal pairs are generated and splits into two parts, including 90% for training and 10% for validation.

### B. Reconstruction of Synthetic Data

We first evaluate the reconstruction performance on the noisy synthetic data with different numbers of exponentials, i.e. spectral peaks, and different sampling rates. These data are not presented in the training dataset and generated following (1) with parameters listed in TABLE I.

The reconstruction errors of the synthetic data are shown in Fig. 5. The region below the red line, denoting a lower reconstruction error than 0.1, is significantly larger for DHMF than for the other three methods. Thus, DHMF can reconstruct signals that have a larger number of exponentials or with fewer samples. Among conventional techniques, the LRHM obtains the best reconstruction. LRHMF is faster than LRHM but shows reconstruction of somewhat lower quality since it is sensitive to an algorithm parameter, which optimal value is not available in practice (See in Section VI. B). The DLNMR does not provide as low reconstruction error as LRHM. With a higher error threshold, e.g. 0.2 marked with white line in Figure 5, LRHM and DLNMR show comparable performance although LRHMF allows reconstructing a slightly larger number of exponentials. Unlike DLNMR, DHMF outperforms LRHMF as it provides low errors and enables more exponentials. For example, at the sampling rate 15%, DHMF enables the reconstruction of 7 exponentials, while LRHMF only allows 5 exponentials. These observations imply that DHMF provides better reconstructions than other methods.

A faithful reconstruction of low-intensity peaks is a challenge in the DL approach [16]. A synthetic signal (Fig. 6(a)), whose spectral intensity of the weakest peak is about 20 times smaller than that of the highest one, is used to evaluate the ability to reconstruct low-intensity peaks. DHMF provides the most consistent spectral peak shape and intensity (Fig. 6(h)) to the fully sampled peak (Fig. 6(d)). DLNMR hardly retrieves the weakest peaks while LRHMF introduces a comparable false peak next to the correct peak. Both LRHM and DHMF reconstruct the two weakest peaks correctly, but the latter provides better reconstruction judged from the higher value of the Pearsons linear correlation coefficients(Fig. 6(c)) defined as:

$$r(\mathbf{a},\mathbf{b}) = \frac{\sum_{i=1}^{n}(\mathbf{a}_i - \overline{\mathbf{a}})(\mathbf{b}_i - \overline{\mathbf{b}})}{\sqrt{\sum_{i=1}^{n}(\mathbf{a}_i - \overline{\mathbf{a}})^2}\sqrt{\sum_{i=1}^{n}(\mathbf{b}_i - \overline{\mathbf{b}})^2}}, \quad (15)$$

where $\overline{\mathbf{a}}$ and $\overline{\mathbf{b}}$ denotes the mean value of the reconstructed and ground-truth spectral intensities $\mathbf{a}$ and $\mathbf{b}$, respectively. The length $n$ for each peak is defined such that the region around the central peak can be almost covered.

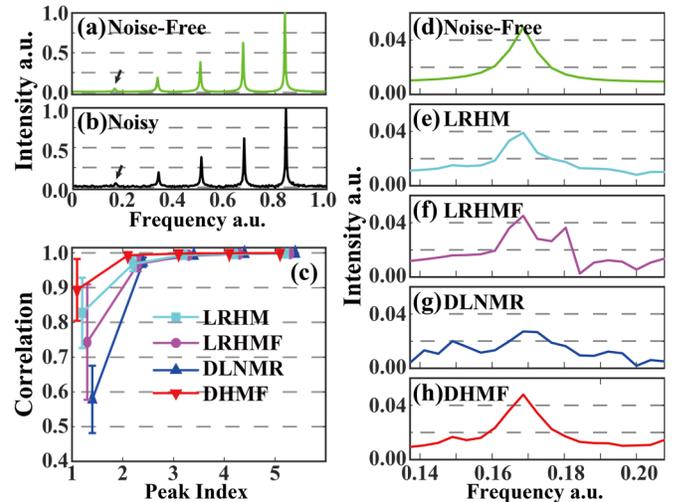

Fig. 6. Synthetic signals with low-intensity peaks reconstructed by low-rank and deep-learning methods. (a) is the fully sampled noise-free signal. (b) is the noisy data with the additive Gaussian noise under a standard deviation of 0.05. (c) is the Pearsons linear correlation coefficient of each peak. (d-h) are the zoomed-in marked weakest peaks of fully sampled noise-free signal and reconstructed signals by LRHM, LRHMF, DLNMR, and DHMF from 25% data, respectively. Note: The average reconstruction error RLNE is computed over 100 sampling trials.

C. *Reconstruction of Realistic NMR Data*

NMR spectroscopy is one of the most powerful tools for the analysis of the composition and structure of chemicals and proteins. The time-domain signal directly acquired by an NMR spectrometer can usually be well modeled as a superposition of damped exponential signals [8-12]. As examples, we present two spectra $^1$H-$^{15}$N TROSY (Fig. 7(a)) and $^1$H-$^{15}$N HSQC (Fig. 8(a)), where we added Gaussian noise (standard deviation $2\times10^{-2}$ and $10^{-2}$ respectively) and performed undersampling with the Poisson gap method [23]. Note that because of the way the two-dimensional NMR experiments are performed, the undersampling is only performed on the first dimension (also called the indirect dimension), and another dimension is commonly fully sampled and not time-consuming[28]. Here, the regularization parameter $\lambda$ and rank parameter $R$ in LRHMF is set to $10^2$ and 10, respectively. The detailed experimental setup is reported in Supplement S2.

The examples in Figures 7 and 8 show that DHMF provides faithful peak recovery, while LRHM may underestimate the intensity of peaks (Fig. 7(g)(h) and Fig. 8(g)(h)), and LRHMF may introduces false peaks and spectral distortions(Fig. 7(d) and Fig. 8(d)). All but DHMF compared methods lose a number of weak peaks indicated by the black circle in Figure 7(c)-(e).

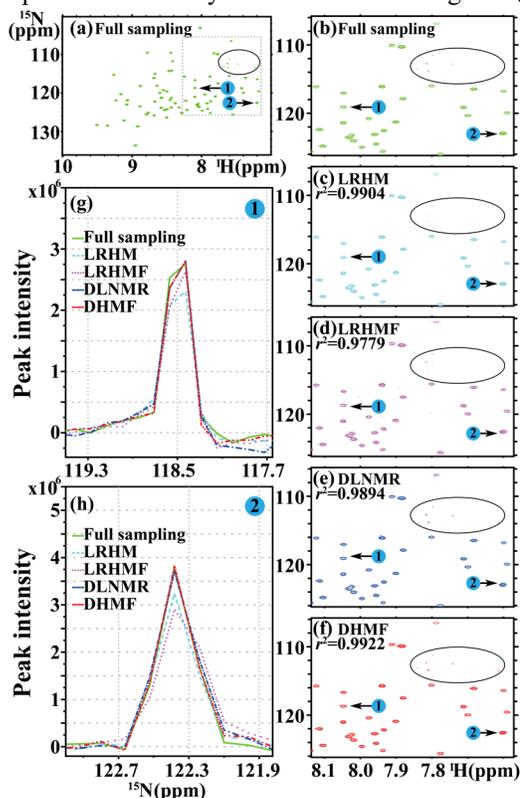

Fig. 7. 2D $^1$H-$^{15}$N TROSY spectrum of Ubiquitin reconstructed by low-rank and deep-learning methods. (a) is the fully sampled NMR spectrum. (b)-(f) are the zoom-in full sampling and reconstructions by LRHM, LRHMF, DLNMR, and DHMF, respectively. (g) and (h) are zoomed in reconstructed 1D $^{15}$N traces. The ppm denotes parts per million by frequency, which is the unit of chemical shift. Note: The data size is 127×512 and undersampling is performed on the dimension of 127 at a sampling rate of 25%.

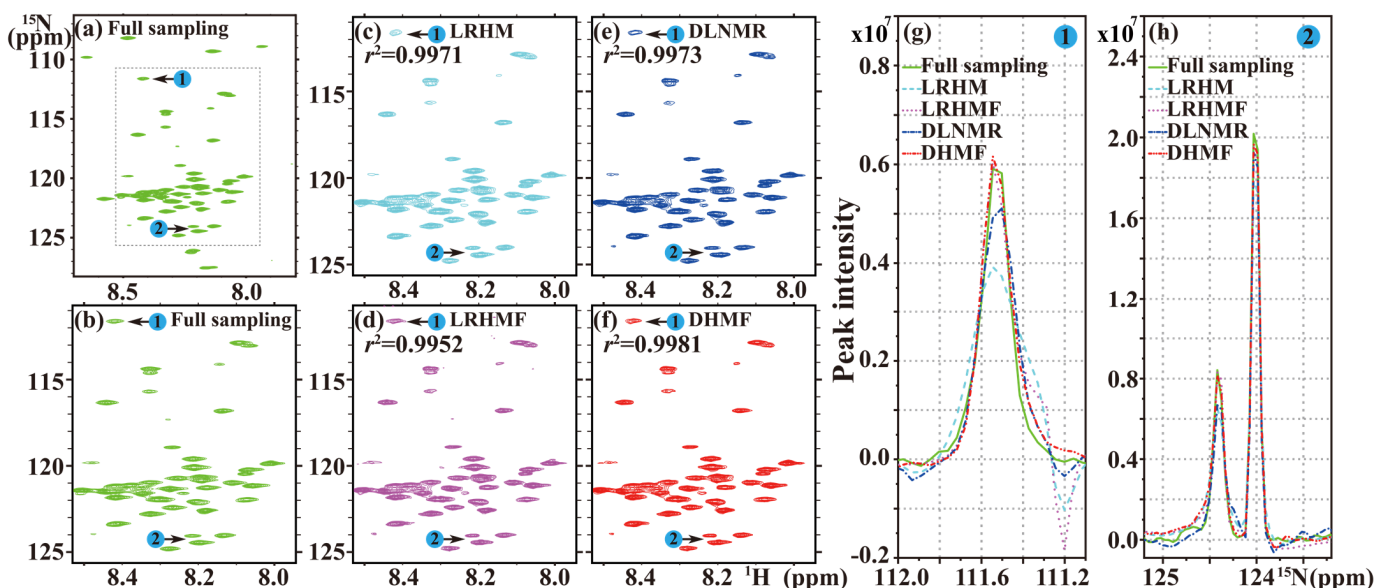

Fig. 8. 2D $^1$H-$^{15}$N HSQC spectrum of protein reconstructed by low-rank and deep-learning methods. (a) is the fully sampled NMR spectrum, (b)-(f) are the zoom-in full sampling and reconstructions by LRHM, LRHMF, DLNMR, and DHMF, respectively. (g) and (h) are zoomed in 1D $^{15}$N traces. Note: The data size is 255×116 and the undersampling is performed on the dimension of 255 at a sampling rate of 25%.

TABLE II
QUANTATIVE SCORES For ESTIMATED PARAMETERS

| Method | Parameters of exponentials | | | |
|---|---|---|---|---|
| | Amplitude | Damping factor | Phase | Frequency |
| LRHM | 2.49±0.99 | 1.94±1.01 | 2.54±1.01 | 2.50±1.01 |
| LRHMF | 2.54±1.01 | 2.45±0.96 | 2.55±1.04 | **2.60±1.05** |
| DLNMR | 2.26±1.12 | 2.63±1.11 | 2.29±1.15 | 2.31±1.14 |
| DHMF | **2.71±1.07** | **2.98±1.04** | **2.62±1.12** | 2.59±1.14 |

Note: Parameter of exponentials are estimated on the reconstructed signals. For each parameter, we provide a score for each method to indicate the relative accuracy among all methods. A method is scored as 4 if it gets the best accuracy, or scored as 3, 2 and 1 if the method takes the second, third and last place among all methods. The number after the notation "±" is the standard deviation over 9000 randomly generated test signals (with different exponential parameters and sampling rates). As an example, we provide the error of estimate parameters for one sampling trail (SUPPLEMENT Fig. S4-1) and the corresponding true parameters (Supplement TABLE S1-3).

The DHMF also shows the highest correlation for the peak intensities (Fig. 7(f) and Fig. 8(f)) among all methods. Therefore, the proposed DHMF method provides the most faithful reconstruction for the realistic NMR spectra.

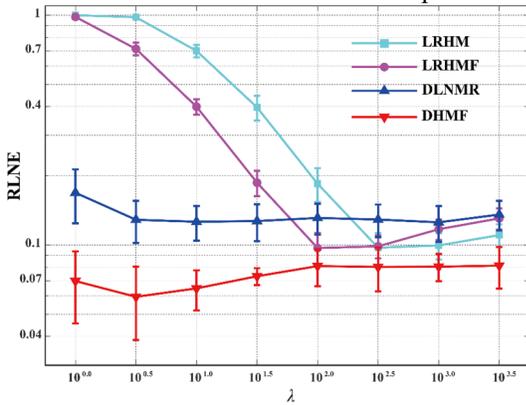

Fig. 9. The regularization parameter $\lambda$ versus reconstruction errors, RLNEs. Note: The sampling rate is 25% and the error bar stands for the standard deviation under 100 resampling trials. Parameters of the synthetic signal are reported in Supplement TABLE S1-2.

## VI. DISCUSSIONS

### A. Estimation of Exponential Parameters

We further evaluate the accuracy of parameters of synthetic exponentials, including frequency, amplitude, damping factor, and phase, that are retrieved from reconstructions by the ESPRIT [58, 59]. A method will score 4 points if it achieves the best accuracy to the ground-truth parameters among all compared methods and scored 1 point if it obtains the worst one. The scores for each parameter are summarized in TABLE II.

The proposed DHMF achieves the highest scores on the amplitude and damping factor and comparable performance on phase and frequency to the best performance of compared methods. Thus, DHMF preserves the amplitude and damping factor much better in the reconstruction.

### B. Robustness of Algorithm Parameters

Discussions on the regularization parameter $\lambda$ for all methods are given here. The regularization parameter $\lambda$ is used for data consistency to maintain the sampled data points. A smaller $\lambda$ corresponds to more noise in the acquired data points. To test the robustness, we measure the reconstruction errors for $\lambda$ value in the range from 1 to $10^{3.5}$. The results shown in Fig. 9 indicate that two deep learning methods DHMF and DLNMR are more robust than the model-based optimization methods LRHM and LRHMF. Besides, the proposed DHMF shows lower reconstructed errors than DLNMR for all regularization parameters.

Selection of parameter $R$ is essential for matrix factorization methods, including LRHMF and DHMF, which determines the size of adjustable variable **P** and **Q**, and the upper bound of the rank of $\mathcal{R}\mathbf{x}$, i.e. the maximal number of exponentials in **x**. Once the true number of exponentials is known, e.g. for synthetic signals, LRHM reconstruction error (Fig. 10(a)) is significantly lower than the parameter is set to a much large value $R$=20 (Fig. 10(c)). In practical cases, since the true $R$ value is not accessible, the performance of LRHM may be suboptimal. On the contrary, DHMF does not require the true number of exponentials. For example, DHMF results obtained with $R$=10 (Fig. 10(b)) and $R$=20 (Fig. 10(d)), differ only slightly. Thus, DHMF is easy to be used in practice.

Results for weight $\gamma$ and step size $\tau$ (Fig. 11(a)(b)) show DHMF has good robustness and reaches lower RLNE with $\gamma$=$10^{-2.0}$ and $\tau$=$10^{-3.0}$ among all compared values. The number of dense layer $L$ and block $K$ decide the number of learnable parameters and influence the reconstruction outcome. A smaller number of learnable parameters, such as $L$ less than 1 and $K$ less than 3, result in RLNE higher than 0.1, whereas a large number of learnable parameters increases demand for the computation memory. For balance, the number of dense layer $L$ and block $K$ of DHMF is set as 8 and 5, respectively.

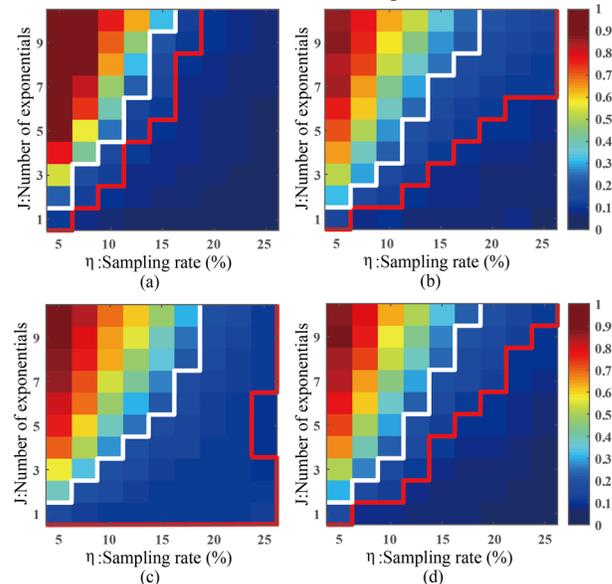

Fig. 10. The average reconstruction errors, RLNEs, for (a)(c) LRHMF with the true rank $R$=10 and empirical setting $R$=20, (b)(d) DHMF with a rank parameter $R$=10 and $R$=20. Note: RLNEs are averaged over 100 Monte Carlo sampling trials.

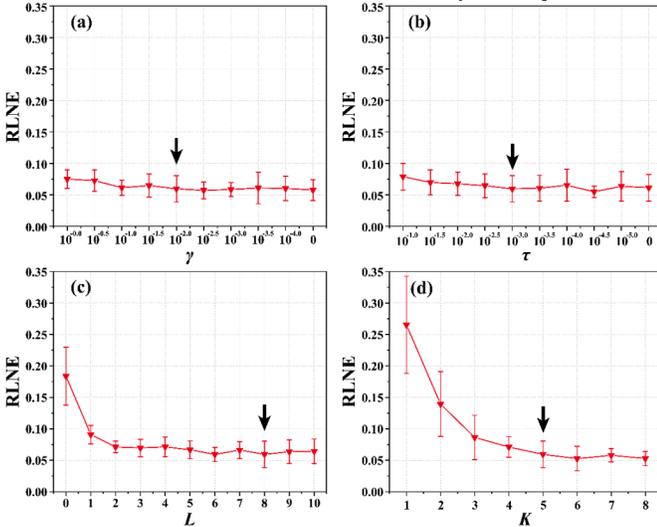

Fig. 11. Reconstruction errors under different values of hyper-parameters. (a) weight $\gamma$ in the loss function. (b) step size $\tau$ in updating module **D**. (c) the number of dense layer $L$. (d) the number of block $K$. Note: The error bars stand for the standard deviation under 100 sampling trials and come from the randomness of sampling pattern. The black arrows denote the chosen value in our paper. Note: The sampling rate is 25%. Parameters of the synthetic signal are reported in Supplement TABLE S1-2.

### C. Architecture of **D** updating module

For **D** updating module, DHMF exactly follows the LRHMF algorithm. We compared this approach with three other strategies described below, including two different CNN versions (named **D**-I and **D**-II) and a version with a learnable step size $\tau$ (named **D**-III).

1) **D**-I: Input of $k$-th **D** updating module is the concatenation of $\mathbf{D}^k$, $\mathcal{R}\hat{\mathbf{x}}^k$ and $\mathbf{P}^{k+1}(\mathbf{Q}^{k+1})^H$:

$$\mathbf{D}^{k+1} = \mathcal{D}_1(\mathbf{D}^k, \mathcal{R}\hat{\mathbf{x}}^k, \mathbf{P}^{k+1}(\mathbf{Q}^{k+1})^H), \quad (16)$$

2) **D**-II: Input of $k$-th **D** updating module is the concatenation of and $\mathbf{D}^k$ and $\mathcal{R}\hat{\mathbf{x}}^k - \mathbf{P}^{k+1}(\mathbf{Q}^{k+1})^H$:

$$\mathbf{D}^{k+1} = \mathcal{D}_2(\mathbf{D}^k, \mathcal{R}\hat{\mathbf{x}}^k - \mathbf{P}^{k+1}(\mathbf{Q}^{k+1})^H), \quad (17)$$

where $\mathcal{D}_1$ and $\mathcal{D}_2$ denote the CNN similar to **P** and **Q** updating module in Fig. 2(c).

3) **D**-III: Except for learnable step size $\tau$, the **D** updating module is the same as the proposed DHMF.

Fig. 12 shows that the proposed DHMF obtains the highest weak peak correlations among all versions.

### D. Network Performance under Mismatch between Trained and Target Signals

Neural networks are always trained for a class of problems that share some common properties and may need to be re-trained when the mismatch of key properties exists between coming problems and training data.

Here, we discuss our network performance under different key properties, including the sampling rate, rank, distribution of spectrum, signal size, distribution of amplitude, and sampling pattern. A problem is defined to be solvable when the reconstruction error, RLNE, is less than 0.1, indicating that the energy loss of the reconstruction is less than 1%.

Fig. 13 shows that, the problems of the sampling rate, rank, distribution of spectrum, signal size, and distribution of amplitude are solvable in a certain range of mismatch. However, the problem of the sampling pattern is hard to be solved without re-training. Detailed experimental setups are provided in Supplement S7, and results are briefly summarized below.

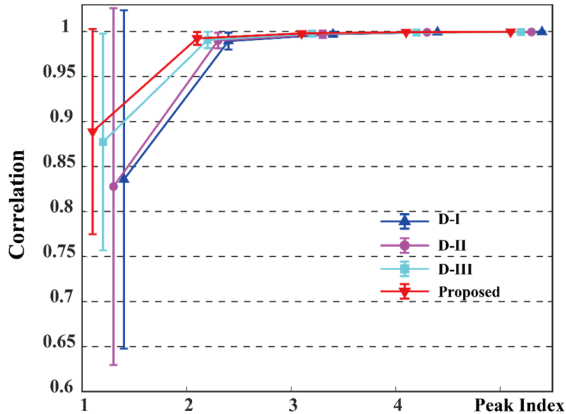

Fig. 12. Pearson linear correlation coefficient of each peak for reconstructing signals (Supplement TABLE S1-2) under 25% sampling rate and 0.05 Gaussian noise level. Correlation are calculated over 100 sampling trails. Mean RLNEs of **D**-I, **D**-II, **D**-III and the proposed DHMF are 0.057, 0.060, 0.072 and 0.054, respectively.

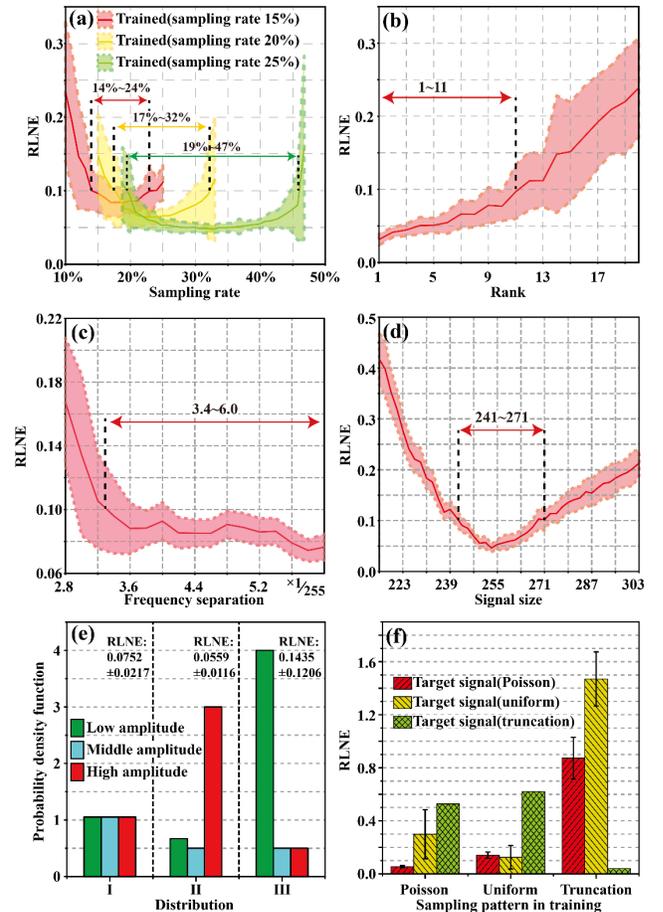

Fig. 13. Reconstruction errors of signal with (a) different sampling rates, (b) rank from 1 to 20, (c) frequency separations of two neighboring peaks from, and (d) size from 215 to 303. (e) is reconstruction errors under three distributions of amplitude. (f) is reconstruction errors for three sampling patterns. Note: Means and the standard deviations of RLNE are computed over 100 sampling trails.

1) Sampling rate (Fig. 13 (a)): The solvable ranges are 14%~24%, 17%~32%, and 19%~47%, when sampling rates of trained signals are 15%, 20% and 25% respectively. Thus, the range of solvable sampling rate increases from narrow to relatively wide with the increase of the sampling rate in trained signals.

2) Rank (Fig. 13 (b)): A target signal is solvable when its rank is no larger than the maximal rank of trained signals. Higher ranks will be unsolvable (RLNE>0.1).

3) Distribution of spectrum (Fig. 13(c)): The distribution of spectrum is defined as the frequency separation of neighboring peaks. The problem is solvable if the frequency separation of the adjacent peak is higher than 3.4/255 Hz. Much closer peaks cannot be reconstructed faithfully.

4) Signal size (Fig. 13 (d)): The trained DHMF is possible to reconstruct the target signal that has a size ±6% of trained signals.

5) Distribution of amplitudes (Fig. 13(e)): It is solvable unless the low amplitude peaks exist in the target signal with high probability.

6) Sampling pattern (Fig. 13(f)): DHMF reconstructs the signal if the training and target sampling belong to the same pattern type, but likely fails if they are different.

### E. Computational Time for Reconstructions

One of the advantages of DL is its low computational complexity. Though the network training, which is performed only once, is time-consuming, the reconstruction runs fast.

TABLE III shows that, for the same hardware (CPU without parallelization), the reconstruction time of DHMF is less than two Hankel-based methods, LRHM and LRHMF, which means much of the reduction in computation time is attributed to the proposed DHMF. DHMF requires more computation time than the state-of-the-art DL method because of the 2D convolution on the Hankel matrix.

Moreover, in common applications, DL methods use GPU acceleration, and two Hankel-based iterative methods use multi-cores CPU acceleration with parallelization. TABLE IV shows that with the acceleration, DL methods are much faster than the two iterative methods.

TABLE III
COMPUTATIONAL TIME WITH SAME HARDWARE
(UNIT: SECOND)

| Size<br>Method | 255 | 511 | 1023 |
|---|---|---|---|
| LRHM | 10.92 | 25.19 | 108.92 |
| LRHMF | 7.06 | 46.17 | 153.79 |
| DLNMR | 0.26 | 0.48 | 0.96 |
| DHMF | 6.06 | 13.91 | 35.74 |

TABLE IV
COMPUTATIONAL TIME WITH HARDWARE ACCELERATION
(UNIT: SECOND)

| Size<br>Method | 255 | 511 | 1023 |
|---|---|---|---|
| LRHM | 1.33 | 5.05 | 40.13 |
| LRHMF | 1.04 | 8.56 | 44.35 |
| DLNMR | 0.03 | 0.04 | 0.06 |
| DHMF | 0.39 | 1.20 | 5.31 |

### F. Comparison with Compressed Sensing

For synthetic exponential signal(Fig. 14) and realistic NMR spectrum(Fig. 15), we compare the DHMF with the state-of-the-art sparsity-based Compressed Sensing (CS) approach [20, 21]. Fig. 14(e) shows that CS has the problem to recover damped exponential signals, especially for broad peaks that lack sparsity. For realistic NMR data, CS underestimates the intensity of these peaks (Fig. R15(e)(f)), while DHMF does not. Besides, the correlation of CS is much lower than those of DHMF. Therefore, the proposed DHMF provides much better reconstruction than CS.

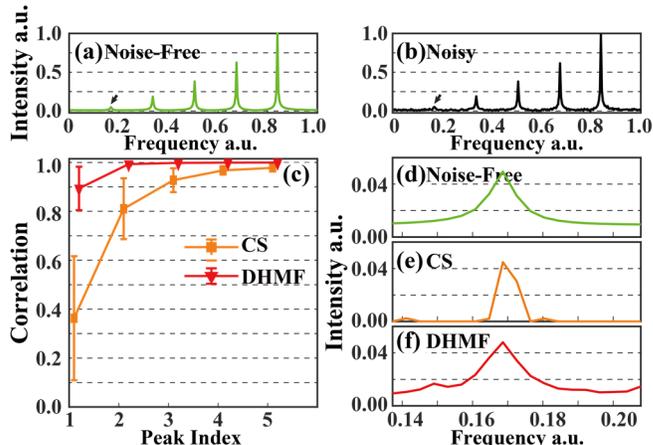

Fig. 14. Synthetic signals with low-intensity peaks reconstructed by compressed sensing and proposed methods. (a) is the fully sampled noise-free signal. (b) is the noisy data with the additive Gaussian noise under a standard deviation of 0.05. (c) is the Pearsons linear correlation coefficient of each peak. (d-f) are the zoomed-in marked weakest peaks of fully sampled noise-free signal and reconstructed signals by CS and DHMF from 25% data, respectively. Note: The average reconstruction error RLNE is computed over 100 sampling trials.

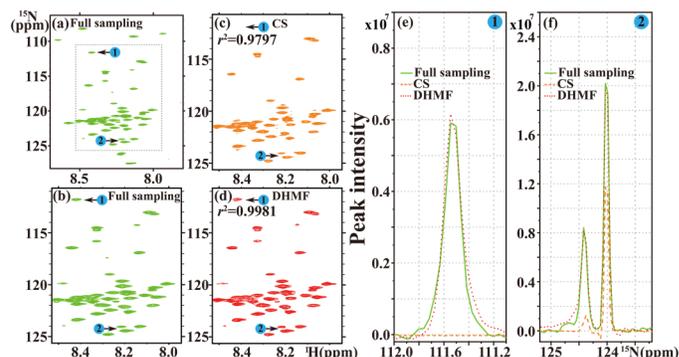

Fig. 15. 2D $^1$H-$^{15}$N HSQC spectrum of protein reconstructed by compressed sensing and proposed methods. (a) is the fully sampled NMR spectrum, (b)-(d) are the zoom-in full sampling and reconstructions from 25% data by CS and DHMF, respectively. (e) and (f) are zoomed in 1D $^{15}$N traces.

## VII. CONCLUSION

In this work, we proposed a new deep learning neural network called DHMF by imitating the model-based matrix factorization for exponential signal recovery. Experimental results on synthetic data and real biological data demonstrate that the DHMF outperforms state-of-the-art model-based and deep learning-based methods on preserving low-intensity signals and obtains a more accurate estimation of the amplitude and damping factor of exponentials. In the future, we will extend the proposed DHMF to other important problems, such as outlier toleration[15].

The shared source can be found on the website of our Computational Sensing Group at Xiamen University.

## VIII. ACKNOWLEDGMENTS

The authors thank China Mobile Group for providing eCloud service and NVIDIA Corporation for GPU donation. The authors also thank Weiping He, Shaorong Fang, and Tianfu Wu from Information and Network Center of Xiamen University for the help with the GPU computing. The authors also thank editors and reviewers for the valuable suggestions and encouragements.

# Supplementary Materials For "Exponential Signal Reconstruction with Deep Hankel Matrix Factorization"

Yihui Huang†, Jinkui Zhao†, Zi Wang, Vladislav Orekhov, Di Guo, Xiaobo Qu*

*S1. Parameters of Synthetic Signals*

TABLE S1-1
SYNTHETIC DATA FOR FIG. 6

| Peaks ID / Parameters | 1 | 2 | 3 | 4 | 5 |
|---|---|---|---|---|---|
| Amplitude ($A$) | 0.100 | 0.300 | 0.500 | 0.700 | 1.000 |
| Damping factor ($\tau$) | 50 | 75 | 100 | 125 | 150 |
| Phase ($\phi$) | 0 | 0 | 0 | 0 | 0 |
| Frequency ($f$) | 0.1655 | 0.3349 | 0.5004 | 0.6698 | 0.8353 |

TABLE S1-2
SYNTHETIC DATA FOR FIG. 3, FIG. 4 AND FIG. 9

| Peaks ID / Parameters | 1 | 2 | 3 | 4 | 5 |
|---|---|---|---|---|---|
| Amplitude ($A$) | 0.5145 | 0.6623 | 0.7253 | 0.7825 | 0.9872 |
| Damping factor ($\tau$) | 26.47 | 35.63 | 48.78 | 61.51 | 81.50 |
| Phase ($\phi$) | $2\pi/5$ | $4\pi/5$ | $6\pi/5$ | $8\pi/5$ | $2\pi$ |
| Frequency ($f$) | 0.1532 | 0.3135 | 0.4716 | 0.6124 | 0.7831 |

TABLE S1-3
SYNTHETIC DATA FOR FIG. S4-1

| Peaks ID / Parameters | 1 | 2 | 3 | 4 |
|---|---|---|---|---|
| Amplitude ($A$) | 0.717 | 1.000 | 0.601 | 0.454 |
| Damping factor ($\tau$) | 173.24 | 126.44 | 31.59 | 107.82 |
| Phase ($\phi$) | 3.5281 | 5.6890 | 2.1928 | 3.8518 |
| Frequency ($f$) | 0.0706 | 0.1534 | 0.4166 | 0.4833 |

*S2. Experiment Setups of NMR Spectra*

NMR spectra we used in the paper include 2D $^1$H-$^{15}$N TROSY spectrum of Ubiquitin protein in Fig. 7 and 2D $^1$H-$^{15}$N HSQC spectrum of CD79b protein in Fig. 8.

The 2D $^1$H-$^{15}$N TROSY spectrum of Ubiquitin was acquired under 298.2K temperature on an 800 MHz Bruker spectrometer. The fully sampled spectrum consists of 127×682 complex points, the direct dimension ($^1$H) has 682 data points while the indirect dimension ($^{15}$N) has 127 data points.

The 2D $^1$H-$^{15}$N HSQC spectrum was acquired for 300 μM $^{15}$N-$^{13}$C labeled sample of cytosolic CD79b in 20 mM sodium phosphate buffer, pH 6.7 at 25 °C on 800 MHz Bruker AVANCE III HD spectrometer equipped with 3 mm TCI cryoprobe. The fully sampled spectrum consists of 1024×255 complex points, the direct dimension ($^1$H) has 1024 data points while the indirect dimension ($^{15}$N) has 255 data points.

*S3. Reconstruction Errors*

TABLE S3-1
AVERAGE RECONSTRUCTION ERRORS (RLNES) OF LRHM

| J \ SR | 5% | 7.5% | 10% | 12.5% | 15% | 17.5% | 20% | 22.5% | 25% |
|---|---|---|---|---|---|---|---|---|---|
| 10 | 0.883 | 0.809 | 0.663 | 0.586 | 0.482 | 0.381 | 0.282 | 0.220 | 0.161 |
| 9 | 0.887 | 0.780 | 0.634 | 0.540 | 0.424 | 0.310 | 0.228 | 0.183 | 0.137 |
| 8 | 0.864 | 0.762 | 0.631 | 0.485 | 0.377 | 0.272 | 0.196 | 0.144 | 0.123 |
| 7 | 0.846 | 0.712 | 0.540 | 0.431 | 0.312 | 0.216 | 0.167 | 0.130 | 0.114 |
| 6 | 0.801 | 0.684 | 0.504 | 0.343 | 0.260 | 0.166 | 0.140 | 0.114 | 0.099 |
| 5 | 0.765 | 0.599 | 0.411 | 0.275 | 0.201 | 0.146 | 0.118 | 0.106 | 0.093 |
| 4 | 0.708 | 0.533 | 0.312 | 0.214 | 0.150 | 0.126 | 0.108 | 0.094 | 0.085 |
| 3 | 0.595 | 0.375 | 0.210 | 0.151 | 0.122 | 0.098 | 0.088 | 0.087 | 0.079 |
| 2 | 0.434 | 0.220 | 0.156 | 0.121 | 0.104 | 0.097 | 0.089 | 0.087 | 0.085 |
| 1 | 0.203 | 0.133 | 0.110 | 0.098 | 0.100 | 0.097 | 0.094 | 0.092 | 0.098 |

TABLE S3-2
AVERAGE RECONSTRUCTION ERRORS (RLNES) OF LRHMF

| J \ SR | 5% | 7.5% | 10% | 12.5% | 15% | 17.5% | 20% | 22.5% | 25% |
|---|---|---|---|---|---|---|---|---|---|
| 10 | 0.896 | 0.833 | 0.691 | 0.609 | 0.476 | 0.309 | 0.184 | 0.141 | 0.109 |
| 9 | 0.904 | 0.793 | 0.653 | 0.554 | 0.388 | 0.245 | 0.165 | 0.127 | 0.104 |
| 8 | 0.876 | 0.769 | 0.635 | 0.472 | 0.337 | 0.207 | 0.148 | 0.118 | 0.102 |
| 7 | 0.854 | 0.718 | 0.535 | 0.416 | 0.260 | 0.176 | 0.137 | 0.112 | 0.104 |
| 6 | 0.802 | 0.691 | 0.472 | 0.301 | 0.217 | 0.140 | 0.121 | 0.107 | 0.096 |
| 5 | 0.763 | 0.589 | 0.376 | 0.245 | 0.173 | 0.135 | 0.116 | 0.108 | 0.096 |
| 4 | 0.694 | 0.504 | 0.277 | 0.187 | 0.135 | 0.126 | 0.111 | 0.101 | 0.095 |
| 3 | 0.565 | 0.331 | 0.186 | 0.140 | 0.120 | 0.106 | 0.101 | 0.100 | 0.093 |
| 2 | 0.380 | 0.187 | 0.145 | 0.122 | 0.110 | 0.108 | 0.104 | 0.103 | 0.102 |
| 1 | 0.153 | 0.115 | 0.107 | 0.101 | 0.108 | 0.107 | 0.106 | 0.104 | 0.112 |

TABLE S3-3
AVERAGE RECONSTRUCTION ERRORS (RLNES) OF DLNMR

| J \ SR | 5% | 7.5% | 10% | 12.5% | 15% | 17.5% | 20% | 22.5% | 25% |
|---|---|---|---|---|---|---|---|---|---|
| 10 | 0.853 | 0.843 | 0.620 | 0.528 | 0.405 | 0.300 | 0.253 | 0.204 | 0.174 |
| 9 | 0.843 | 0.735 | 0.579 | 0.482 | 0.350 | 0.268 | 0.215 | 0.179 | 0.151 |
| 8 | 0.824 | 0.701 | 0.558 | 0.414 | 0.317 | 0.240 | 0.190 | 0.152 | 0.138 |
| 7 | 0.808 | 0.648 | 0.484 | 0.362 | 0.263 | 0.199 | 0.173 | 0.141 | 0.125 |
| 6 | 0.750 | 0.610 | 0.424 | 0.305 | 0.242 | 0.165 | 0.154 | 0.133 | 0.113 |
| 5 | 0.695 | 0.519 | 0.346 | 0.252 | 0.197 | 0.157 | 0.140 | 0.124 | 0.105 |
| 4 | 0.648 | 0.451 | 0.294 | 0.221 | 0.169 | 0.138 | 0.130 | 0.112 | 0.105 |
| 3 | 0.529 | 0.359 | 0.216 | 0.170 | 0.138 | 0.121 | 0.113 | 0.111 | 0.103 |
| 2 | 0.390 | 0.231 | 0.169 | 0.144 | 0.128 | 0.115 | 0.115 | 0.109 | 0.108 |
| 1 | 0.190 | 0.133 | 0.134 | 0.125 | 0.134 | 0.117 | 0.118 | 0.108 | 0.119 |

TABLE S3-4
AVERAGE RECONSTRUCTION ERRORS (RLNES) OF DHMF

| J \ SR | 5% | 7.5% | 10% | 12.5% | 15% | 17.5% | 20% | 22.5% | 25% |
|---|---|---|---|---|---|---|---|---|---|
| 10 | 0.874 | 0.890 | 0.632 | 0.504 | 0.332 | 0.224 | 0.189 | 0.127 | 0.107 |
| 9 | 0.890 | 0.759 | 0.573 | 0.440 | 0.262 | 0.196 | 0.148 | 0.114 | 0.093 |
| 8 | 0.844 | 0.722 | 0.525 | 0.377 | 0.233 | 0.162 | 0.134 | 0.096 | 0.087 |
| 7 | 0.836 | 0.651 | 0.446 | 0.285 | 0.181 | 0.138 | 0.116 | 0.087 | 0.078 |
| 6 | 0.781 | 0.602 | 0.366 | 0.228 | 0.164 | 0.104 | 0.095 | 0.079 | 0.067 |
| 5 | 0.719 | 0.491 | 0.294 | 0.181 | 0.120 | 0.098 | 0.087 | 0.076 | 0.061 |
| 4 | 0.659 | 0.390 | 0.222 | 0.140 | 0.100 | 0.085 | 0.076 | 0.065 | 0.057 |
| 3 | 0.503 | 0.259 | 0.150 | 0.104 | 0.078 | 0.070 | 0.064 | 0.059 | 0.050 |
| 2 | 0.299 | 0.146 | 0.117 | 0.081 | 0.070 | 0.065 | 0.060 | 0.053 | 0.051 |
| 1 | 0.138 | 0.078 | 0.076 | 0.065 | 0.063 | 0.059 | 0.058 | 0.049 | 0.046 |

*S4. Parameter Estimation of Random Synthetic Signals*

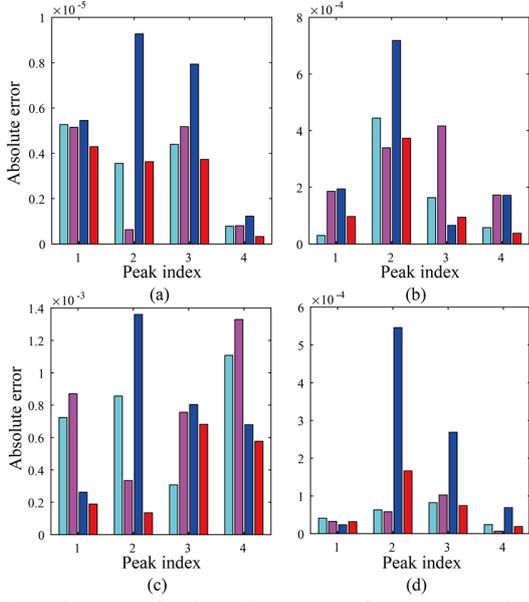

Fig. S4-1. The error of estimated parameters from reconstructions of one random undersampling. (a)-(d) denote the errors of estimated frequency, amplitude, damping factor and phase, respectively. The cyan, purple, blue, and red color correspond to the LRHM, LRHMF, DLNMR, and DHMF method, respectively. Note: Ture parameters of the signal are listed in TABLE S1-3.

*S5. Measurement of Finding Reconstructed Peaks*

Peak $\hat{\mathcal{T}}_1$ is considered as a reconstruction for the ideal peak $\mathcal{T}_1$ since it is in the range of $d_{max}$, while reconstruction for the ideal peak $\mathcal{T}_2$ is missing since all reconstructed peaks are out of the range $d_{max}$.

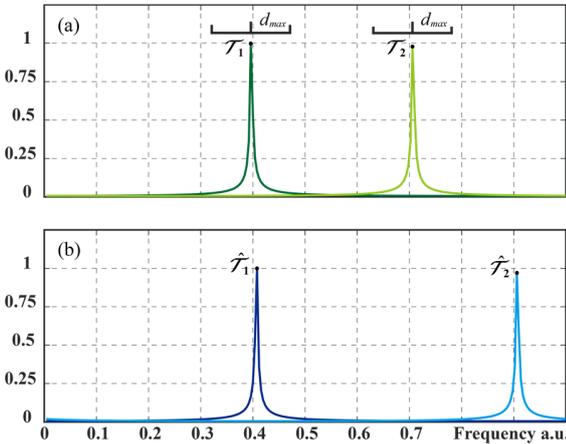

Fig. S5-1. An illustration of the measurement of finding the reconstructed peak to ideal one. dmax is the maximal range that one can find a reconstructed peak to the ideal one.

*S6. Singular Values with Logarithmic Axis*

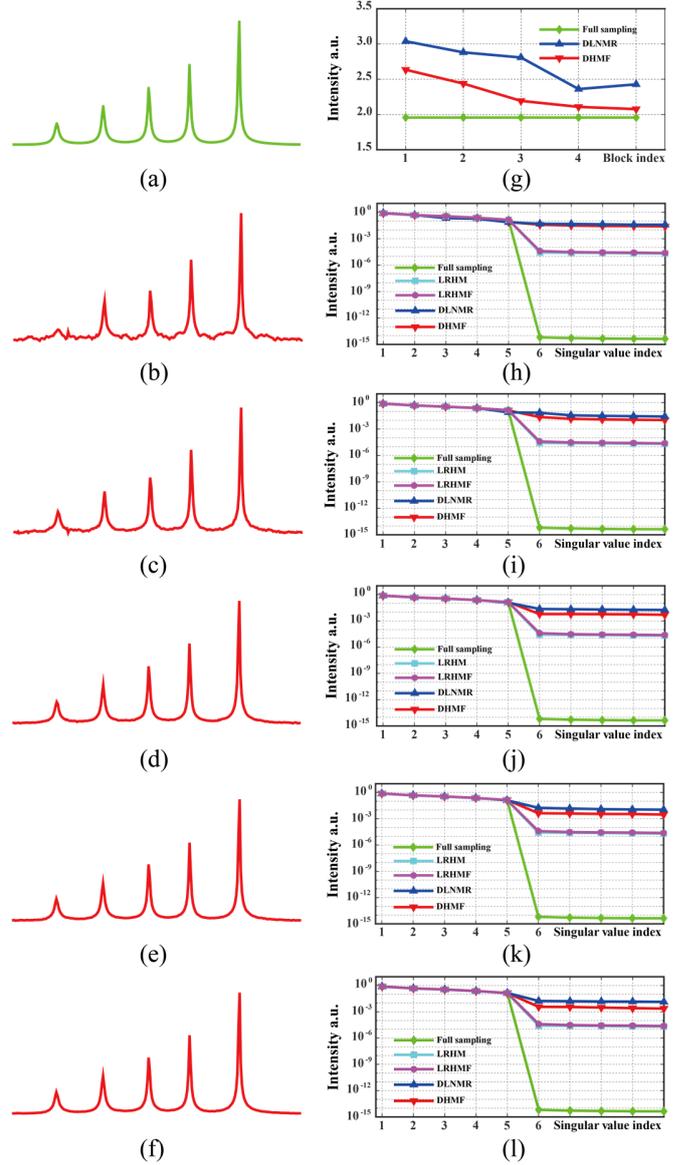

Fig. S6-1. The reconstructed spectra and singular values at each block with Logarithmic Axis. (a) fully sampled spectrum, (b)-(f) the reconstructed spectrum by the 1st to 5th blocks, (g) the nuclear norm of Hankel matrix of time domain signal, and (h)-(l) denote corresponding singular values of the ouput of each block.

*S7. Network Performance under Mismatch between Trained and Target Signals*

A. Distribution of the spectrum:

In Fig. 13(c), the synthetic signals with five peaks are generated with the same damping factor, amplitude and phase, except for the frequency separation. Solid line denotes mean, along with shaded area marking the standard deviation, over 100 sampling trails under sampling rate 25%. The shaded area comes from the randomness of sampling pattern.

B. Signal size:

In Fig. 13(d), the synthetic signals with five peaks have same parameters as Supplement TABLE S1-2. Solid line denotes mean, along with shaded area marking the standard deviation, over 100 sampling trails of signal under sampling rate 25%. The shaded area comes from the randomness of sampling pattern.

C. Distribution of amplitude

In Fig. 13(e), the synthetic signals with five peaks are generated with the same frequency (0.1, 0.3, 0.5, 0.7, 0.9), damping factor 50, and zero phase, while the amplitude is from 0.05 to 1.0 and is chosen by corresponding distribution independently.

The distribution of amplitude for trained signals is the uniform distribution. The range of low, middle and high amplitude is [0.05,0.2), [0.2 to 0.8] and [0.8 to 1.0], respectively. The range of [0.00,0.05) is treated as noise since the noise level is 0.05 (the standard deviation of noise).

In Fig. 13(e), distribution I is uniform distribution, distribution II has more peaks with high amplitude, and distribution III has more peaks with low amplitude.

D. Sampling pattern:

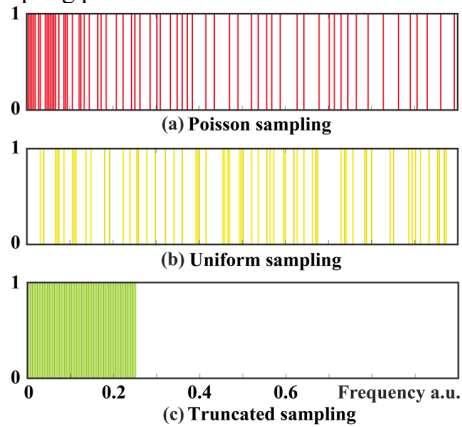

Fig. S7-1. Three common sampling patterns. (a)-(c) are patterns of Possion, uniform and truncation. Note: The error bars in Fig. 13(f) stand for the standard deviation under 100 sampling trails of signal (Supplement TABLE S1-2) at sampling rate 25% and randomness does not exist in truncation pattern.

*S8. Network Performance on Outlier Toleration*

Outlier toleration [1] is one of the study topics about low rank Hankel matrix. Here, we discuss the outlier toleration of DHMF without/with retraining the network. Notably, the network architecture of DHMF has not changed.

In our experiments, the outliers uniformly occur and the real and imaginary parts of additive corruption are drawn uniformly from the interval $[-c\mathbb{E}(\text{Re}(x)), c\mathbb{E}(\text{Re}(x))]$ and $[-c\mathbb{E}(\text{Im}(x)), c\mathbb{E}(\text{Im}(x))]$, respectively, where Re() and Im() denote the real and imaginary part and $\mathbb{E}()$ is the expectation. Constant c is set as 4 (same to [1]).

Without re-training the network, the proposed DHMF cannot reconstruct the signal with outliers since the reconstruction error is larger than 0.1 (Fig. S8-1(d)) and the spectrum has obvious distortions (Figs. S8-1 (e)-(f)).

With re-training the network from new signals that are contaminated by outliers, the DHMF can handle the reconstruction with outliers under very low rate of outliers (10% in Fig. S8-2 (d)). This performance is much worse than other advanced method[1] which can handle the outlier rate up to 50%.

Thus, the DHMF have very limited ability to handle outliers. This observation is reasonable since outlier problems are commonly solved by adding $l_1$-norm in the reconstruction[1], rather than the $l_2$-norm loss defined in our method. A possible approach may be incorporating the $l_1$-norm in the loss function of network and re-train the network. This task is beyond the scope of this paper and we will explore it in the future work.

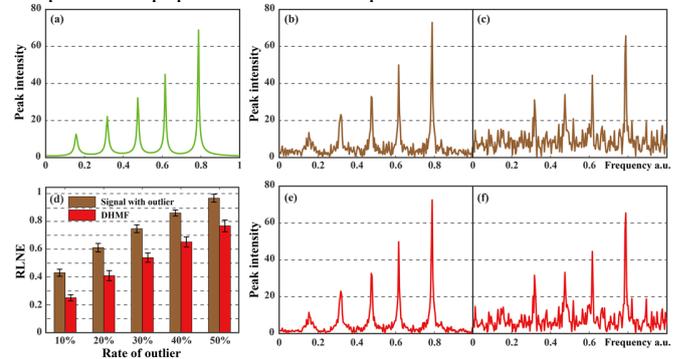

Fig. S8-1. Reconstruction results of DHMF for outliers without retraining. (a) Noise-free full sampling. The spectra with (b) 10% and (c) 50% rate of outlier and (e)(f) are the reconstructed spectra, respectively. (d) RLNE of corrupted signals and reconstructed signals by DHMF under different rate of outlier. Note: The error bars stand for the standard deviation under 100 resampling trails of signals and come from the randomness of location and intensity of outlier.

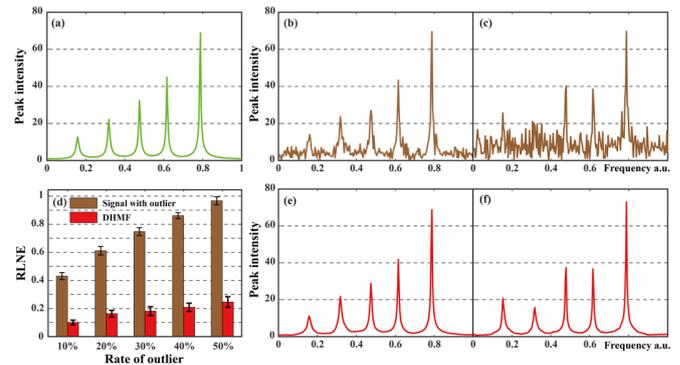

Fig. S8-2. Reconstruction results of DHMF for outliers. (a) Noise-free full sampling. The spectra with (b) 10% and (c) 50% rate of outlier and (e)(f) are the reconstructed spectra, respectively. (d) RLNE of corrupted signals and reconstructed signals by DHMF under different rate of outlier.